\RequirePackage{fix-cm}
\documentclass{svjour3}                     
\smartqed  

\usepackage{graphicx}
\usepackage{color}

\begin{document}

\title{Modelling gravitational waves from precessing black-hole binaries: 
Progress, challenges and prospects
}

\author{Mark Hannam
}

\institute{Mark Hannam \at
              School of Physics and Astronomy \\
              Cardiff University \\
              The Parade \\
              Cardiff CF24 3AA \\
              UK \\
              \email{mark.hannam@astro.cf.ac.uk}        
}

\date{Received: date / Accepted: date}

\maketitle

\begin{abstract}
The inspiral and merger of two orbiting black holes is among the most promising sources for the first
(hopefully imminent) direct detection of gravitational waves (GWs), and measurements of these signals
could provide a wealth of information about astrophysics, fundamental physics and cosmology. 
Detection and measurement require a theoretical description of the GW signals from all possible
black-hole-binary configurations, which can include complicated precession effects due to the 
black-hole spins. Modelling the GW signal from generic precessing binaries is therefore one of the most
urgent theoretical challenges facing GW astronomy. This article briefly reviews the phenomenology
of generic-binary dynamics and waveforms, and recent advances in modelling them.

\end{abstract}

\section{Introduction}
\label{intro}

Our first major physical theory was Newton's law of gravitation, which was powerful enough to describe
both the fall of fruit on Earth and the orbits of planets in the solar system. Newton's calculations of
 two-body dynamics from one general law were the central triumph of the scientific revolution. 
 Now Newton's theory of gravitation has been upgraded to Einstein's general theory of 
relativity, where two-body dynamics can become far more complex, and may once again be the
catalyst for discoveries about the universe. The motion of two orbiting bodies has been modified in
two important ways. (1) All acceleration generates gravitational waves; those generated by the
orbital motion carry away energy, which causes the orbital separation to slowly decrease, and the bodies
to spiral together. (2) The angular momenta of the bodies are no longer individually conserved, and
in general their individual angular momenta (spins), and the orbital plane of the binary, all precess. 
These effects are weak when the 
two bodies are far apart (being only a small correction to the Newtonian description), but for compact 
objects like neutron stars and black holes, which may continue to orbit up to very 
small separations, the precession effects can lead to wild complex dynamics, before the inspiral terminates
in a collision that produces an intense final burst of gravitational radiation. 

The extreme physics of  black-hole and neutron-star inspiral and merger 
may soon be measurable by the new 
gravitational-wave observatories Advanced LIGO (aLIGO) and Advanced Virgo (AdV), which are due
to come online in 2015-16~\cite{Harry:2010zz,aVIRGO}. 
The primary goal of these experiments is to make the first direct
detection of gravitational waves, and our current understanding of astrophysics suggests that 
``compact binary coalescences''  are most likely to provide the first
detections; indeed, when aLIGO and AdV reach their design sensitivity around 
2018~\cite{Aasi:2013wya}, they
may observe neutron-star and black-hole mergers on a daily basis~\cite{Abadie:2010cf}. 

Our ability to observe the gravitational waves from binary mergers, and in particular to measure the
binary's properties from the GW signals (their masses and angular momenta, and the binary's location
in the sky and distance from Earth), all depend on theoretical predictions of the signals produced by all
possible binary configurations. 

The GW signal from the slow inspiral can be predicted by a post-Newtonian (PN) expansion of the Einstein
equations~\cite{Blanchet:2013haa}. 
The late inspiral and merger requires numerical solutions of the full nonlinear Einstein 
equations~\cite{Centrella:2010mx}. Both approaches are necessary because the PN approximation 
breaks down as the 
black holes approach merger, and numerical-relativity (NR) simulations are too computationally expensive
to generate waveforms that cover the thousands of orbits that could be observable by aLIGO and AdV.
The construction of complete inspiral-merger-ringdown (IMR) models for
generic-binary configurations in principle require NR simulations across a seven-dimensional
parameter space. By 2010 models had been produced for a subset of binary configurations that do
not precess (the black-hole spins are parallel to the binary's orbital angular momentum, as I will 
describe later). In these
configurations the waveforms have a simple structure, and most of the signal power resides in the 
dominant waveform harmonic, and we need only model its (relatively simple) amplitude and phase.
These models required 20--30 numerical 
simulations~\cite{Ajith:2009bn,Pan:2009wj,Santamaria:2010yb,Taracchini:2012ig,Taracchini:2013rva}. 
A simple counting argument suggests that extending these
models to generic configurations would require tens of thousands of simulations. In addition, the 
precession spreads the signal power across other waveform harmonics, making the modeling
of these waveforms (not just describing a single waveform, but also the subtle waveform variations
with respect to binary configurations) far more challenging. In the strong-field regime near merger 
the detailed phenomenology of generic binaries can be determined only after the numerical simulations 
have been performed, so a judicious sampling of that parameter space is difficult to estimate a priori. 
For these reasons, modeling the GW signal from generic binary mergers has become the most urgent 
theoretical challenge facing gravitational-wave astronomy.

The statements above summarize our understanding of the problem around 2010. Since then a number of important
studies have shown how to factorize out the complex precession effects from the waveforms, and have
indicated that we can, to a good approximation, produce generic models by applying a time-dependent rotation
to non-precessing-binary waveforms, based on the precessional dynamics of the particular configuration we
wish to describe. In other words, we can ``twist up'' the non-precessing-binary models that we already have
to produce generic models. This
is an approximation that does not remove all of the issues in generic-binary waveform modeling, and so far
includes only a rudimentary treatment of the merger and ringdown phase --- but these results nonetheless mark
a huge step forward in the modeling of generic systems, and suggest that a solution sufficient for the needs
of gravitational-wave astronomy may be possible in the near future. 

The purpose of this review is to expand on the above. The focus is on the goal of constructing IMR models
for generic-binary systems. I will summarize the basic phenomenology of generic-binary dynamics and
waveforms (Sec.~\ref{sec:spin}), the PN description of the inspiral (Sec.~\ref{sec:inspiral}), and then IMR
models from the current non-precessing-binary models (Sec.~\ref{sec:NP}) through to the latest results
in generic-binary modelling (Sec.~\ref{sec:generic}). At the end I will discuss some of the remaining issues
and challenges (Sec.~\ref{discussion}).

\section{The effect of spin on binary dynamics and waveforms}
\label{sec:spin}

The effect of spin on the dynamics of two compact bodies was first considered by 
Barker and O'Connell in 1975~\cite{Barker:1975ae}. 
Extensive reviews of early work are given in Refs.~\cite{Barker:1979,Damour87}. 
The phenomenology of the binary dynamics and the gravitational-wave
signals are discussed and illustrated in detail by Apostolatos, {\it et. al.}~\cite{Apostolatos:1994mx} and 
Kidder~\cite{Kidder:1995zr}. In this section I summarize these effects in the order of their impact on the 
GW signal, which mimics the progress of GW-modelling efforts in recent years. 

In the following we consider two black holes of masses $m_1$ and $m_2$, and refer to the total mass
as $M = m_1 + m_2$. The mass-ratio between the two black holes is $q = m_1 / m_2$, where 
$m_1 > m_2$ and $q > 1$, although other conventions are also used in the literature. A less ambiguous
indicator of the mass ratio is $\eta = m_1 m_2 / M^2$, which is independent of whether we define 
$q > 1$ or $q < 1$. If the black-hole spin-angular-momenta are $\mathbf{S}_1$ and $\mathbf{S}_2$, 
then the dimensionless spins are $\chi_1 = |\mathbf{S}_1|/m_1^2$ and 
$\chi_2 = |\mathbf{S}_2| / m_2^2$, and $\chi_i \in [0,1]$, where $\chi = 1$ 
corresponds to an extreme Kerr black hole. 

We consider only black-hole binaries following non-eccentric inspiral, because we expect that by the time
most binaries enter the aLIGO and AdV sensitivity bands, any eccentricity present at the binary's formation will
have essentially radiated away~\cite{Peters:1964zz}.

\subsection{Non-precessing systems}

\begin{figure*}
  \includegraphics[width=0.32\textwidth]{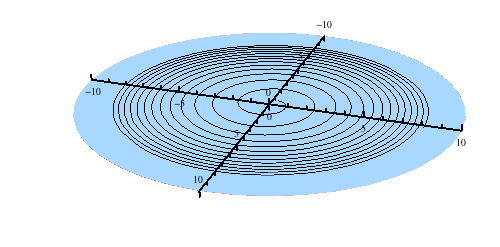}
    \includegraphics[width=0.32\textwidth]{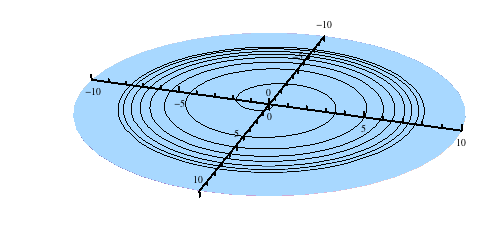}
      \includegraphics[width=0.32\textwidth]{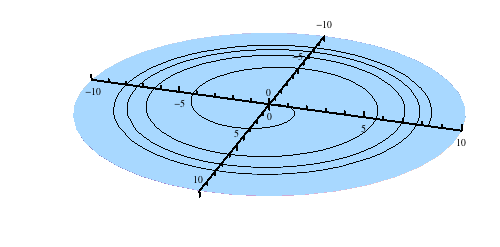}
\caption{Phenomenology of inspiral for aligned-spin systems. The figures show the motion of one black
hole in a mass-ratio 1:4 binary, from a separation of $D \approx 8M$ through to merger. The central figure depicts
a binary made up of nonspinning black holes; in the left figure the spins are parallel to the orbital angular momentum, 
and in the right figure they are in the opposite direction; the spins are $\chi_i = 0.75$. 
The effect of the spin on the inspiral rate is clear.}
\label{fig:q4}    
\end{figure*}

If two black holes do not spin, or if their spins are parallel to the orbital angular momentum of the binary,
then the direction of the orbital plane of the binary is fixed. 
The black holes follow non-eccentric, but slowly decaying orbits. 
The rate of decay is determined by the loss of energy due to gravitational radiation, and this is a function both 
of the black-hole masses, but also their spins. In a post-Newtonian expansion of the equations of motion,
we can see that if the spins are parallel to the orbital angular momentum, then the rate of energy loss is reduced,
and the black holes inspiral more slowly, e.g., see Eqn.~(1) in Ref.~\cite{Buonanno:2002fy}. 
If the spins are in the opposite direction to the orbital angular momentum,
then the black holes inspiral more quickly. This effect is illustrated in Figure~\ref{fig:q4} for three spin
configurations of a mass-ratio 1:4 binary. As the inspiral proceeds, angular 
momentum is radiated from 
the system and the orbital angular momentum of the binary decreases; although the spin magnitudes can vary 
in some PN treatments (see, for example, Ref.~\cite{Blanchet:2006gy}), in all waveform models discussed here 
the spin magnitudes are treated as constant.

\subsection{precession}

\begin{figure}
  \includegraphics[width=0.9\textwidth]{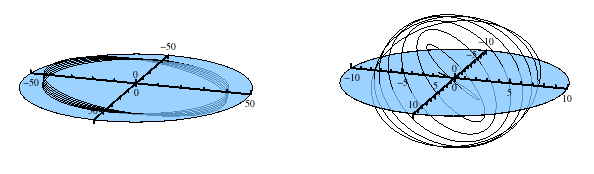}
\caption{As in Fig.~\ref{fig:q4}, but for a mass-ratio 1:3 binary, where the larger black hole's spin is 
$\chi_1 = 0.75$,
and is perpendicular to the orbital angular momentum, leading to significant precession. The precession 
over several orbits is mild at large separations (left), but leads to complicated dynamics near merger (right).}
\label{fig:precession}      
\end{figure}

When the spins are not aligned with the orbital angular momentum, then spin-orbit and spin-spin couplings
lead to precession of the spins and orbital plane. This is a purely relativistic effect: in Newtonian physics the
individual angular momenta of the two bodies, and of the total system, are all individually conserved. 
In the absence of gravitational radiation, the direction of 
the {\it total} angular momentum $\hat{\mathbf{L}}$ would remain
fixed, and the orbital and spin angular momenta all precess around it, i.e., $\dot{\mathbf{J}} = 0$ and so 
$\dot{\mathbf{L}} = - \dot{\mathbf{S}}$. For single-spin systems it is possible to estimate the rate of precession, i.e., 
the angular speed of $\hat{\mathbf{L}}$ about $\hat{\mathbf{J}}$, to be $\Omega_p \propto J / r^3$, where $r$ is the 
separation of the binary~\cite{Apostolatos:1994mx}. When we include radiation-reaction effects, we find that in most
cases the direction of $\hat{\mathbf{J}}$  remains approximately fixed; it is certainly the closest we have to fixed
direction during the inspiral. Figure~\ref{fig:precession} shows an example of the orbital motion of 
one of the black holes in a precessing binary.

The exceptions to this description are configurations where the black-hole spins are {\it almost} anti-aligned
with the total angular momentum, such that $|\mathbf{J}| = | \mathbf{L} + \mathbf{S} |\approx 0$. 
Now there is no fixed precession axis,
and the orbital plane ``tumbles'' in space~\cite{Apostolatos:1994mx}. 
Loss of angular momentum through gravitational-wave emission means that this situation cannot persist
indefinitely; $|L|$ decreases while $|S_1|$ and $|S_2|$ remain fixed, and $|J|$ does not remain small. 
For this reason the phenomena is known as transitional precession. Although a fascinating effect, transitional
precession will be rare in binaries that are observable by GW detectors: it requires very particular 
spin configurations, which would have to be met while the binary is within the detectors' sensitivity band.

\subsection{Waveforms} 
\label{sec:waveforms}

We can see the effect of precession on the gravitational-wave signal simply by considering the leading-order
quadrupole contribution, which takes into account only variations in the moment of inertia of the binary, i.e., the
accelerations of the two bodies. For a binary in orbit, the wave signal estimated from the  quadrupole 
approximation is directed predominantly perpendicular to the orbital plane. This is why, in a non-precessing 
binary, the GW signal is strongest directly above or below the plane of the binary. This dominant contribution to the 
signal can be represented entirely by the $(l=2, |m|=2)$ spin-weight $(-2)$ spherical harmonics, which means 
that the signal will be the same for all orientations of the binary, up to an overall amplitude factor and phase offset. 
This simplifies both
the modeling of these signals --- to a good approximation we can focus on only the $(2,\pm2)$ modes --- and 
searches for them in detector data. Most searches comb through the data with a template bank of theoretical
waveforms, and in these non-precessing configurations the search template bank does not need to include 
waveforms that vary with respect to binary orientation and detector polarization, because these do not 
change the functional form of the
waveform; the search only need cover a parameter space of the two masses and spin magnitudes. 
(To date actual searches
in detector data have considered only nonspinning black holes, so the template banks have been two-dimensional;
the most recent examples are Refs.~\cite{Colaboration:2011np,Aasi:2012rja}).

When the orbital plane precesses, the principal emission direction will also precess. Now the waveform 
structure is more complicated. If the inclination angle between $\mathbf{L}$ and $\mathbf{J}$ is small, 
and the wave signal is decomposed with respect to spin-weighted spherical harmonics 
$^{-2}Y_{\ell m} (\theta, \phi)$ that are defined such that the $\theta = 0$ direction (the $\hat{z}$ axis)
is aligned with $\mathbf{J}$, then the signal can again be represented to a good approximation by 
only the $(2, \pm2)$ modes. But there will now be a quadrupole contribution to other $l=2$ harmonics, which grows 
as the $\hat{\mathbf{L}}$-$\hat{\mathbf{J}}$ inclination angle is increased. Furthermore, the signal will now 
vary significantly depending 
on the relative orientation of the binary to the detector; now that the dominant quadrupole contribution 
to the signal is made up of all $l = 2$ harmonics, the 
variations in the waveform with binary orientation cannot be treated as an overall amplitude factor in a search
(although of course the orientation-dependence of the waveform {\it can} be described analytically, by varying 
$(\theta,\phi)$ in the spherical harmonics). 

\begin{figure}
  \includegraphics[width=1.0\textwidth]{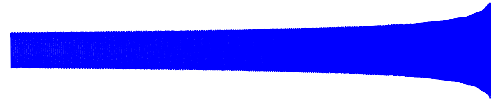}
  
    \includegraphics[width=1.0\textwidth]{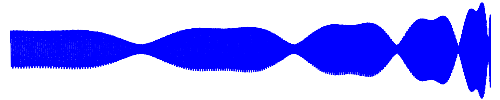}
\caption{Gravitational waveforms for the mass-ratio 1:3 system shown in Fig.~\ref{fig:precession}. In the upper
figure the observer is oriented approximately along the direction $\hat{\mathbf{J}}$, 
and the effect of precession
on the waveform is minimal. In the lower figure the observer is perpendicular to $\hat{\mathbf{J}}$,
and significant modulation effects are visible.}
\label{fig:PrecessingWaveforms}       
\end{figure}

This is illustrated in Fig.~\ref{fig:PrecessingWaveforms}, which shows the GW signal from a precessing 
mass-ratio 1:3 binary from two different orientations, where the observer is aligned with $\hat{\mathbf{J}}$,
and where the observer is perpendicular to $\hat{\mathbf{J}}$. We might loosely refer to these orientations
as ``face-on'' and ``edge-on'', but this is misleading. Since $\hat{\mathbf{L}}$ precesses around $\hat{\mathbf{J}}$,
the {\it average} direction of the normal to the orbital plane lies approximately along $\hat{\mathbf{J}}$, but the
observer is in fact {\it never} exactly face-on to the binary. Similarly, in configurations with a large 
$\hat{\mathbf{L}}$-$\hat{\mathbf{J}}$ inclination angle,
we may choose orientations where the binary alternates between being face-on and edge-on as the 
precession progresses.
Nonetheless, the waveform viewed from the $\hat{\mathbf{J}}$-aligned  
orientation appears very similar to one from a non-precessing binary, and indeed would be difficult to 
distinguish in a GW observation. It is also important to note that it is the waveform phase, not amplitude, that
has the dominant effect in waveform measurements, and these figures are a meaningful illustration of the
differences in the waveforms with respect to orientation only because modulations in waveform amplitude and
in phase are closely related. If that were not the case and these two waveforms had the same phase evolution, 
these two waveforms would be effectively equivalent if used in a search template bank.

Having summarized the phenomenology of binary dynamics and waveforms, I will move on to the details of
waveform modelling.

\section{In the beginning: inspiral}
\label{sec:inspiral} 

The early inspiral of the binary, from which thousands of GW cycles may be visible to ground-based 
detectors, can be described by a post-Newtonian (PN) approximation to the Einstein equations. 
A detailed summary of PN calculations of the orbital dynamics and waveforms is beyond
the scope of this short review; the reader is referred to Blanchet~\cite{Blanchet:2013haa} for a thorough review of the current
status of PN calculations, and
to Apostolatos {\it et. al.}~\cite{Apostolatos:1994mx} and Kidder~\cite{Kidder:1995zr} for detailed illustrations of the 
phenomenology of precessing systems. 

The key ingredients in the construction of most PN waveforms for generic systems following adiabatic
quasicircular inspiral are: (1) The orbital frequency
evolution, which depends on the total mass $M$, the mass ratio $\eta$, and the components of the 
bodies' spins $\mathbf{S}_i$.
(2) The precessional dynamics, expressed as equations of motion for $\mathbf{L}$ and $\mathbf{S}_i$. 
(3) The waveform polarizations as a function of the orbital motion. 

Ingredients (1) and (2) arise from the equations of motion for the conservative dynamics (i.e., without gravitational
radiation), plus the gravitational-wave flux terms. 
In the equations of motion the non-spinning terms are known up to 3.5PN 
order~\cite{Blanchet:2013haa}, and the energy for circular orbits up to 4PN~\cite{Jaranowski:2013lca}.  
Spin-orbit coupling terms are also known up to 3.5PN order~\cite{Hartung:2010jg,Marsat:2012fn}.
Spin-spin couplings are known to next-to-leading-order (NLO, 3PN), and some terms are also known to
next-to-next-to-leading-order (NNLO, 4PN)~\cite{Levi:2011eq,Hartung:2011ea}. 
In the flux NNLO (3.5PN) spin-orbit terms are known~\cite{Bohe:2013cla}, and ``tail'' effects to next-to leading order 
(4PN)~\cite{Marsat:2013caa}, while spin-spin effects are known only to leading order (2PN)~\cite{Gergely:1999pd}.

Note that the expressions for the GW phasing at a given PN order are not unique, and that depending on
the way the PN terms are truncated in the calculation of a given ``approximant'', 
the results will differ in the error terms at higher PN orders. 

The GW polarization amplitudes are known to 3PN order in nonspinning terms~\cite{Blanchet:2013haa}, 
although the 
3.5PN contribution to the $(\ell=2, m=2)$ harmonic is also known~\cite{Faye:2012xt}. 
The spin terms are known only to their respective (spin-orbit, spin-spin, and tail) leading orders, which
means that the highest PN order for which all spin terms are consistently known is 2PN~\cite{Arun:2008kb}. 

Note that some of these terms were calculated only in the last year. Some higher-order spin terms are likely 
to appear in the near future, while others (for example calculations to 4PN order in the flux) 
will be far more challenging. The convergence
properties of the PN expansion are poorly understood, and it is not clear how many more terms are 
necessary to reach the desired accuracy for GW observations. Our current understanding, based on 
comparisons of all of the available approximants, is that nonspinning PN waveforms are sufficiently accurate
for low-mass systems where the merger and ringdown are at the high-frequency edge of the detectors'
sensitivity band (i.e., for detection of neutron-star binaries~\cite{Buonanno:2009zt}), but that the uncertainty in 
spinning PN waveforms from low-mass systems (in particular neutron-star--black-hole binaries) is large
enough to cause some reduction in detection efficiency~\cite{Nitz:2013mxa,Brown:2012qf}. 
The accuracy of PN waveforms for many non-precessing-binary configurations has been quantified 
near merger (where the PN waveforms are {\it least} accurate) by comparison with fully general-relativistic 
NR calculations~\cite{Hannam:2007ik,Boyle:2007ft,Hannam:2007wf,Hannam:2010ec,MacDonald:2012mp}, 
and these comparisons have provided independent validation for the use of 
PN waveforms during the inspiral. 

The accuracy of the PN waveforms could in principle be improved by an appropriate resummation of
the series to improve its convergence properties. This has been achieved to a remarkable degree by 
the effective-one-body (EOB) programme. The conservative dynamics 
are mapped to the 
motion of a test particle in an effective metric, which is the Schwarzschild metric deformed by the 
symmetric mass ratio $\eta$. This was first proposed by Buonanno and Damour in 1999~\cite{Buonanno:1998gg}.
In follow-up work the flux terms responsible
for radiation reaction were also resummed (this time by Pad\'e resummation) 
to produce the
full inspiral dynamics, and the GW signal was constructed at leading (quadrupole, or ``restricted'') PN 
order~\cite{Buonanno:2000ef}. The inspiral waveform was then matched to a ringdown waveform 
at the ``light ring''
(an unstable orbit for massless particles), motivated by test-mass and close-limit-approximation 
results. Those results constituted the first prediction of a full inspiral-merger-ringdown (IMR) waveform.

The first EOB model for non-spinning binaries in 2000 was followed in 2001 by an extension to 
spinning objects~\cite{Damour:2001tu}, using an effective Kerr metric, although it was later found 
 that similar accuracy
can be obtained by simply augmenting the non-spinning EOB Hamiltonian with standard PN
spin terms~\cite{Buonanno:2005xu}. 

These models were based on heuristic arguments for the phenomenology of the merger (in particular
that the inspiral makes a rapid transition to ringdown at a certain point, and this point can be estimated
as the light ring), but had to wait for full numerical simulations in 
2005~\cite{Pretorius:2005gq,Campanelli:2005dd,Baker:2005vv} 
for confirmation and further extension.

\section{Start simple: complete models for non-precessing binaries}
\label{sec:NP}

The first IMR model calibrated to numerical simulations was produced
using a different (``phenomenological'') approach~\cite{Ajith:2007qp}, motivated by the need in GW searches 
for computationally efficient
frequency-domain models. This work proposed separate phenomenological ansatzes for the GW phase 
and amplitude, based on PN results for the inspiral, empirical observations of the late-inspiral/merger 
phase~\cite{Buonanno:2006ui}, and ringdown results from perturbation theory. The coefficients in these 
ansatzes were determined from NR
simulations of nonspinning binaries, and these in turn were used to produce an analytic fit of the coefficients
across the parameter space. One of the key results of this work (besides the construction of the first
NR-based IMR model), was that the phenomenological coefficients varied almost linearly with respect
to the binary's mass ratio, suggesting that a complete nonspinning-binary model could be constructed from 
only a small number ($\sim$5) of NR simulations. 

Spinning binaries present a far greater challenge, with six additional parameters (the components of the
two spin vectors). Modelling efforts began first with the simple subset of non-precessing binaries, where
the spins are parallel or anti-parallel to the binary's orbital angular momentum. As described earlier, the
dominant effect of the spin in these cases is to increase or decrease the rate of inspiral; it will also 
modify the spin of the final black hole, since to first approximation $J_{\rm final} = L + S_1 + S_2$, where
we consider the orbital and spin angular momenta just prior to merger~\cite{Buonanno:2007sv};
more accurate expressions of the final spin based on NR simulations have been calculated for
both non-precessing~\cite{Hemberger:2013hsa} and generic binaries~\cite{Barausse:2009uz}. 
Besides these spin effects, the basic structure of the
waveforms remains unchanged from nonspinning configurations, which greatly simplifies the modelling.

In PN theory we also find that the dominant spin effect on the inspiral rate arises from a combination 
(essentially a weighted sum) of the two spins~\cite{Poisson:1995ef}. Ajith, {\it et. al.} realized that this allows 
us to construct a simple non-precessing-binary model with only three physical parameters 
$(M, \eta, \chi_{\rm eff})$~\cite{Ajith:2009bn,Santamaria:2010yb}, 
where $\chi_{\rm eff}$ is an total-effective-spin, $\chi_{\rm eff} = (m_1 \chi_1 + m_2 \chi_2)/M$, which is 
closely related to the ``reduced spin'' combination that appears in PN 
expressions~\cite{Poisson:1995ef,Ajith:2011ec}.
Such reductions in the parameter 
space are important: they identify the dominant physical effects in the waveforms, and therefore the
physical parameters (or, more often, and unfortunately, {\it combinations} of physical parameters) that
could be measured in GW observations; they reduce the number of templates necessary in a GW 
search, which is crucial to make such searches computationally feasible; and they potentially reduce
the parameter space of necessary NR simulations to construct models. 
The total-effective-spin was used in the first NR-based non-precessing-binary IMR model, which was
proposed in Ref.~\cite{Ajith:2009bn} (using 26 NR simulations) with only minor modifications to the 
procedure introduced in Ref.~\cite{Ajith:2007kx}, and later refined to use the most accurate 
PN expressions for the 
inspiral~\cite{Santamaria:2010yb}. 

In the case of the EOB IMR models, it was found that the EOB dynamics, while they display 
much-improved convergence properties over the
original PN calculations, do not agree sufficiently at the known PN order with fully general-relativistic
NR results near merger. 
(Note that while the phenomenological models require the PN waveforms to be accurate up to 
only $\sim$10 orbits before merger, when they make a transition to an NR-calibrated model, EOB
models require the EOB dynamics to be accurate all the way to merger.) To overcome
this, a ``pseudo-4PN'' term was introduced and calibrated to equal-mass nonspinning 
NR simulations~\cite{Buonanno:2007pf}. Further adjustable parameters were later introduced into 
the EOB Hamiltonian, the flux terms, the resummation of the waveform modes, and (when extending to 
spinning systems) the matching time to ringdown modes; all of these parameters were fit to NR simulations. 

Nonspinning ``EOBNR'' models were introduced shortly after the first phenomenological model, 
and successively refined with additional NR simulations, further adjustable parameters, and a more
accurate resummation (factorization) of the waveform modes~\cite{Damour:2008gu,Pan:2010hz};
the most recent nonspinning-binary model is presented in 
Ref.~\cite{Damour:2012ky}, and a model with harmonics beyond $\ell=2$ is presented in Ref.~\cite{Pan:2011gk}.
A first non-precessing-binary model was introduced in 2010~\cite{Pan:2009wj}.
Unlike the phenomenological models, it included the magnitudes of both black-hole spins.
It was updated in 2012~\cite{Taracchini:2012ig} 
based on an improved spinning EOB Hamiltonian~\cite{Barausse:2009xi}, and more recently calibrated to a 
larger set of 27 numerical simulations~\cite{Taracchini:2013rva}.

\section{The final challenge: generic binaries} 
\label{sec:generic}

The leap from non-precessing to generic binaries is large: the phenomenology
of the binary dynamics and waveforms becomes far more complex; in addition, 
the parameter space grows from three (or two) intrinsic parameters, to seven. Let us enumerate the 
parameters more clearly. A number of parameters of the GW signal can be modified (or added) 
analytically to a single waveform, and so do not need to be included in an underlying model, or 
in numerical simulations to construct the model. These include the total mass of the system, 
which is an overall scale factor in the waveforms; the orientation of the binary with respect to the detector,
or the orientation of a detector with respect to the source on the sky; and the distance of the source from 
the detector.
We may also trivially apply a time shift to a waveform, and an overall physical rotation to the entire binary. 
(For nonprecessing binaries, this rotation is equivalent to a constant phase offset in the waveform.)
We are now left to model the waveforms with respect to the binary's mass ratio, and the three components
of each black hole's spin. Note that, although there may be approximate degeneracies between these
parameters, the essential parameter space remains formally seven-dimensional in non-eccentric
binaries. 

Several approximate degeneracies between configurations may simplify the modelling problem, but it 
is important to emphasize that they are only approximate, and in particular may break down completely 
through merger and ringdown. We've already seen one example, the total effective spin $\chi_{\rm eff}$ 
in the phenomenological non-precessing-binary models. In the ringdown it is instead the final total spin
that characterizes the binary, and so the validity of the effective total spin in parametrizing the waveforms 
weakens in higher-mass binaries, where the detector is more sensitive to the merger and 
ringdown~\cite{Purrer:2013ojf}.

Another example is an approximate degeneracy between 
cases where the components of the black-hole spins in the orbital plane have the same {\it relative}
orientation, i.e., the same angle between the in-plane spin components: if we rotate both spins in the 
orbital plane, to leading order this does not change the waveform
beyond an overall phase change that can be incorporated analytically. This near-equivalence has been
exploited in the past in PN models (see, for example, Ref.~\cite{Buonanno:2002fy}), but it is not an 
exact degeneracy because a rotation of the spins 
and a phase change do not commute. An easy way to see this is to consider  
the ``superkick'' configuration~\cite{Gonzalez:2007hi,Campanelli:2007cga,Brugmann:2007zj}. 
The black holes have equal spins, both in the orbital plane, and 
oppositely directed, such that the orbital plane does not precess, but does bob up and down due to the
emission of linear momentum out of the plane. When this effect terminates at merger, 
the final black hole recoils out of the orbital plane. The magnitude and direction (up or down) of the 
recoil will vary with the initial direction of the spin vectors in the plane. The final recoil is an unambiguous 
physical effect (it is independent of the binary and observer orientations) that depends on the initial spin 
direction, and cannot be removed by a mere phase change. During inspiral the effect of this spin angle 
on the waveform is minimal, and can be ignored to a good approximation, but not during merger and 
ringdown.

In order to develop a general procedure to model generic systems through merger and ringdown, 
we are faced with two problems. The first is to choose
the configurations to simulate numerically. If we make the naive estimate that, based on the non-precessing-binary
phenomenological models, we need $\sim$4 simulations in each direction of the parameter space, then for our
seven-dimensional parameter space, we will need on the order of tens of thousands of simulations. 
We do not expect this to be computationally feasible in the next few years. 
The second problem is that, given these waveforms,
we need a means to model the far more complex waveform structure of precessing systems. Prior to 2010, it was
not clear how to solve either of these problems. A preliminary effort was made to bridge the gap between 
generic inspiral PN waveforms and the ringdown~\cite{Sturani:2010yv}, but that model's physical fidelity  
was not tested beyond a small number of equal-mass binaries, and it has not lead to a 
general approach to generic-binary modelling.

\subsection{Quadrupole alignment}

The key to solving the second problem --- the complicated mode structure and amplitude and phase modulations
of precessing-binary waveforms --- lies in the qualitative description of waveforms in Sec.~\ref{sec:waveforms}. 
The dominant GW emission directions are perpendicular to the plane of the binary, and in a coordinate system
aligned with that direction, most of the signal's power resides in the $(\ell=2, |m|=2)$ spin-weighted spherical
harmonics. Although a full description of the GW signal is more complicated, we nonetheless expect that, so
long as our coordinate system is appropriately aligned with respect to the orbital plane, then the GW signal
will be relatively simple. In other words, if we describe the waveform in a ``co-precessing'' coordinate system,
then generic-binary waveforms will take on a far simpler form than in a real observer's inertial frame. 

A co-precessing frame was first used by Buonanno, Chen and Valisneri~\cite{Buonanno:2002fy} in the study 
of post-Newtonian inspiral waveforms. They observed that the waveforms were indeed simpler in this frame;
in fact, the waveform amplitude and phase modulations were removed. This observation lead them to extend 
the representation proposed in Ref.~\cite{Apostolatos:1995pj} of a generic waveform as a simple 
``carrier waveform'' modulated by two time-dependent polarization tensors.  
Ultimately, though, the co-precessing frame was used primarily as a technical tool in the generation of 
PN waveforms, in particular later in Ref.~\cite{Pan:2003qt}, and to motivate a model that 
includes a carrier wave plus modulations~\cite{Buonanno:2002fy}. 

Eight years later, a co-precessing frame arose again as a consequence of the work of Schmidt, 
{\it et. al.}~\cite{Schmidt:2010it}. Their original goal was to track the precession of a binary from the 
gravitational-wave signal alone, since time-delay effects make it difficult in numerical simulations 
to match the orbital precession at the binary source to the corresponding effects in the waveform at the
observer. They noted that, since the dominant GW power lies in the quadrupole $(\ell=2, |m| = 2)$ modes 
when the coordinate
system is appropriately aligned, it follows that the instantaneous direction of the orbital angular momentum 
could be identified with the orientation of the coordinate system for which the $(\ell=2,|m|=2)$ modes of the 
waveform were maximized.  In doing so, the precession can be tracked from the {\it GW signal alone}.
It turns out that this direction is {\it not} exactly normal to the orbital plane, but in fact follows the direction 
of the orbital angular momentum; in general the two do not coincide, as can be seen by a simple analysis of a 
PN expansion of the orbital angular momentum. They referred to this accelerating frame of reference as the 
``quadrupole aligned'' (QA) frame. 

An alternative co-precessing frame was proposed in 2011~\cite{O'Shaughnessy:2011fx}, defined by the principal 
axes of the GW signal. It was later shown in Ref.~\cite{Boyle:2011gg} that this reduces to the 
QA frame if the principal-axis calculation is restricted to only the $l = 2$ harmonics. 
The main purpose of Ref.~\cite{Boyle:2011gg} was to complete the construction of a unique co-precessing frame 
with a third rotation. This corresponds to a time-dependent phase shift in the co-precessing waveform, 
and its importance is clear in examples of long (PN) waveforms, where the additional phase shift is 
necessary to recover a smooth monotonic frequency evolution in the co-precessing waveform.

\subsection{Mapping to non-precessing binaries}

The most important consequence of the QA frame was
the observation that the waveforms in the QA
frame not only have the same simple (non-modulated) form as non-precessing-binary waveforms, 
but that the \emph{entire mode structure} of the corresponding non-precessing-binary waveform seems to have 
been reproduced. 

The identification between quadrupole-aligned (QA) and non-precessing-binary waveforms 
was observed in Ref.~\cite{Schmidt:2010it} for the mode amplitudes. As we have discussed previously,
it is not the waveform amplitude, but the phase, which is most important in GW observations and
measurements. But, remarkably, Schmidt, {\it et. al.} later found that a correspondence between each 
QA waveform and a non-precessing-binary counterpart could also be extended to the phase, and therefore to
the entire waveform~\cite{Schmidt:2012rh}. They saw that, to a good approximation, the inspiral part of a
QA waveform is the
{\it same} as the waveform produced by a non-precessing system with the {\it same values} of the 
non-precessing spin components. In fact, since the motivation was to 
extend the single-effective-spin 
phenomenological models to precessing systems, the authors tested their claim against systems with 
only the same value of the total effective spin $\chi_{\rm eff}$ defined from the aligned-spin components,
and so the identification was between systems with the same values of $(M, \eta, \chi_{\rm eff})$.

They also noted a crucial corollary. Let us split the spins into those components parallel to the
orbital angular momentum, $\chi_{||}$, and those perpendicular to the orbital angular momentum, i.e., 
approximately in the plane of the binary, $\chi_{\perp}$. If we can factorize a 
precessing-binary waveform $h(M, \eta, \chi_{||}, \chi_\perp)$
into a non-precessing-binary waveform $h(M, \eta, \chi_{||})$ (where the common parameters are the 
same in both 
waveforms), plus a rotation defined by the precession dynamics, then the modelling problem has been 
reduced to finding a model for the precession angles alone. Given such a model, we can use it to construct any 
generic-binary waveform out of its underlying non-precessing-binary waveform. The efficacy of this approach
is demonstrated for one example in Ref.~\cite{Schmidt:2012rh}, and this prescription is 
 is effectively applied in Ref.~\cite{Lundgren:2013jla} to produce a closed-form frequency-domain 
 generic-binary inspiral model.

This identification is not
expected to hold through ringdown, because the ringdown waveform is determined by the final
black hole's spin, and this depends more strongly on the in-plane components of the individual black-hole
spins prior to merger. One configuration was studied in Ref.~\cite{Schmidt:2012rh}, and it was found that 
the identification seems to hold up until the beginning of the ringdown. 
Pekowsky, {\it et. al.}~\cite{Pekowsky:2013ska} examined merger waveforms in more detail.
They showed that the QA merger-ringdown waveform once again agrees well with a non-precessing 
counterpart, but, consistent with Ref.~\cite{Schmidt:2012rh}, there is no simple identification between the 
QA waveform and its non-precessing counterpart, as there is in the inspiral regime. This lead them to 
conclude that IMR models based on twisting up non-precessing waveforms may not be sufficiently 
accurate for GW applications.

\subsection{Twisting waveforms} 

The caveats of Pekowsky {\it et. al.} notwithstanding, generic-binary models now {\it have} 
been produced by twisting up non-precessing waveforms. 

The ``PhenomP'' model~\cite{Hannam:2013oca} provides a simple implementation of the 
procedure proposed in Ref.~\cite{Schmidt:2012rh}. This model uses one of the phenomenological models as the 
underlying non-precessing-binary model. The model of
the precessional dynamics is provided by PN results; these are closed-form 
frequency-domain expressions for the inclination angle between the orbital and total angular 
momenta, and the precession angle, augmented by recent NNLO calculations~\cite{Bohe:2013cla}. 
The ringdown
non-precessing-binary model is modified based on predictions of the final black hole's spin~\cite{Barausse:2009uz}.
Finally, the authors claim that they can adequately model generic precessing waveforms using only {\it three}
physical parameters. The symmetric mass ratio and effective total spin (parallel to the orbital angular 
momentum) carry over from the non-precessing-binary model, and follow from the 
results in Ref.~\cite{Schmidt:2012rh}. For
the remaining four in-plane spin components, they exploit the observation that the dominant precession 
effects can be parameterized by only {\it one} ``precession spin'' parameter, $\chi_p$, which is effectively 
an average of the relevant PN spin terms through the inspiral~\cite{Schmidt:2013aa}. 
With only two spin parameters, they 
choose to write the final model in terms of the spin of only one of the black holes. That this is a reasonable
choice (if not necessarily optimal; studies are ongoing) follows from the arguments above, and also from the
results in~\cite{Pan:2003qt}, which show that the additional modulations in the dynamics in two-spin systems will 
be minor and in most cases not observable in gravitational-wave observations. 
The initial orientation of the spin in the plane
is treated as an overall factor, which is the approximate degeneracy discussed in detail at the beginning of this 
section. 

PhenomP is a proof-of-principle model that involves many approximations: 
that the single-effective-spin 
approximation carries over to the precessional motion, that the precession effects can be parametrized by a
single ``precession spin'' parameter $\chi_p$, that the stationary phase approximation (used to translate to the frequency 
domain) can be made through merger and 
ringdown, and that the PN expressions for the precession angles can be continued through merger and 
ringdown. The first two approximations are motivated by prior work, but the last two have no justification
a priori. Nonetheless, 
the model performed extremely well when compared against a number of hybrid PN-NR waveforms, 
which included two-spin configurations and configurations with a high degree of precession. 

The proposal of twisting non-precessing-binary waveforms has also been adopted to construct a generic 
EOB model~\cite{Pan:2013rra}. Here the non-precessing-binary 
``SEOBNR'' model is used as the underlying model. The precessional dynamics are produced by solving 
the EOB equations of motion for the precessing system. The reason for twisting up the underlying 
non-precessing-binary model, rather than calculating the waveform directly from the EOB dynamics 
using the results of~\cite{Arun:2008kb}, is that the more accurate EOB-factorized waveform modes are
known only in the non-precessing case. 
The resulting inspiral precession waveform is then connected to the ringdown waveform
associated with the correct final spin, which is estimated, as in Ref.~\cite{Hannam:2013oca} 
by the empirical fits in Ref.~\cite{Barausse:2009uz}. The procedure to construct the inspiral precession 
waveform again relies on the observation of Ref.~\cite{Schmidt:2012rh} that each precessing configuration 
has a specific non-precessing counterpart. As with the non-precessing-binary EOBNR model,  this model 
does not make any reduction of the number of physical parameters: it uses all six spin components. 

Beyond the approximations used to generate the inspiral waveforms, which are all well-motivated and
backed up by studies of PN waveforms, the main open issue in these models is the treatment of the 
merger and ringdown. The results of Ref.~\cite{O'Shaughnessy:2012ay} suggest that precession effects continue
through the ringdown. The precession angles continue to evolve in PhenomP, but using only a naive
continuation of the PN expressions used during the inspiral. As with typical frequency-domain PN 
expressions for the waveform phase, the expressions 
for the precession angles cannot be expected to be accurate through merger, but they do remain physically 
reasonable, and their accuracy does not have as strong an effect on the waveform as the phase. 
However, although these expressions appear to perform well in
the comparisons in Ref.~\cite{Hannam:2013oca},  it should
be noted that in low-mass binaries most of the signal power is due to the inspiral, and so errors in the 
merger and ringdown have little effect on the overall agreement of two waveforms, while in high-mass
binaries, where only the merger and ringdown will be detectable, there are very few GW cycles in the detectors'
sensitivity band, and it is much easier to achieve a strong agreement after exploiting our freedom to make
arbitrary relative time and phase shifts.  

In contrast to the PhenomP model, in the EOBNR model no precessional transformation is 
performed during the ringdown. We know from Ref.~\cite{O'Shaughnessy:2012ay} that this is also incorrect, 
but, very likely for the same reasons as
just argued above, the EOB model agrees well when compared against
the two NR waveforms used for comparison in Ref.~\cite{Pan:2013rra}.

\section{Challenges and prospects}
\label{discussion}

There has been tremendous progress in the last few years in the modeling of gravitational waveforms 
from generic binaries. Until recently an accurate model of the full inspiral, merger and ringdown 
from generic binaries was considered a challenge that may not be met in time for the first GW detections. 
Now that has changed, and just in the last year two IMR models of generic binaries were 
proposed~\cite{Hannam:2013oca,Pan:2013rra}. 

Nonetheless, a number of questions and challenges remain. 

The current IMR models incorporate minimal physical information about the effects of precession on
the merger and ringdown. It is not clear how difficult it will be to model those effects --- can we again
extract the main features from a subspace of the full binary parameter space, and if so, how does this
subspace differ from that during the inspiral? In doing this, we must address the question of the 
required length of NR simulations. In non-precessing systems estimates of the required NR waveform 
lengths ranged from hundreds of
orbits before merger if one wishes to eliminate all systematic bias from 
observations~\cite{MacDonald:2011ne,Damour:2010zb,Boyle:2011dy}, down to 
tens of orbits if we allow small parameter biases at levels that are not expected  to have any astrophysical
impact~\cite{Hannam:2010ky,Ohme:2011zm}. It seems unlikely that long waveforms 
will be required to model the precession effects near merger, but no studies have yet addressed these
questions. We must also ask how much of the merger physics will be detectable in GW observations. 
Although the merger is in the strong-field regime of Einstein's theory, the variations in the waveforms
with respect to spin configurations are likely to be subtle and potentially invisible to ground-based
detectors. It is important to clarify these issues, and prioritize the physical effects that need to be modeled. 

We should also note that the current generic IMR models consider only the $\ell = 2$ harmonics 
of the waveforms.
Although these are likely to be sufficient for detection, at least for comparable-mass binaries 
(for the non-precessing-binary case, see Ref.~\cite{Capano:2013raa}), the effect on parameter 
estimation is yet to be fully understood. 

In determining the necessary physical fidelity of the models, the impact on parameter 
estimation is likely to be the deciding factor. This also includes an understanding of the limits of parameter
estimation inherent in the approximate degeneracies of the physical configurations. In 
non-precessing systems, we know that such a degeneracy between the mass ratio and spin will
limit our ability to accurately measure the binary 
masses~\cite{Cutler:1994ys,Baird:2012cu,Hannam:2013uu,Ohme:2013nsa}. 
The efficacy of an effective-total-spin
in the phenomenological~\cite{Ajith:2009bn,Santamaria:2010yb} and PN~\cite{Ajith:2011ec}
models tells us that it will be difficult to measure
individual spins~\cite{Purrer:2013ojf}. And our ability to parameterize precession effects with a single precession spin
in Ref.~\cite{Hannam:2013oca} suggests that relative spin orientations will be difficult to determine. The details of these
degeneracies need to be understood, not just to clarify the possibilities and limitations of astrophysical
measurements from GW observations, but also to inform the regions of binary parameter space that
most urgently need to be modeled. 

All of these degeneracies are approximate, and can be disentangled in sufficiently strong signals. But
we should bear in mind that the accuracy requirements of the theoretical 
models depends on the strength of the signals. By definition most observations with aLIGO and AdV
will be close to 
the detection threshold of signal-to-noise ratio (SNR) $\sim$10, and since no GW observations were made 
with the previous generation of detectors (which were sensitive to $\sim$100--1000 times less volume of
the universe), SNRs above $\sim$30 are unlikely.

Associated with all of these issues is the question of how many NR waveforms are needed, and where
in the parameter space they must come from. The answer to that question will likely emerge as part of
the process of building waveform models: we won't know which waveforms we need until we can be
sure that we don't need any more. 
Meanwhile, the production of NR waveforms proceeds at an encouraging
pace. As striking examples, in the last year a collection of 224 generic waveforms were used in the study
in Ref.~\cite{Pekowsky:2013ska}, and a catalog of 171 generic waveforms was presented in 
Ref.~\cite{Mroue:2013xna}. In addition, the Numerical Injection Analysis (NINJA) 
collaboration~\cite{Aylott:2009ya} has produced a catalog of 64 PN-NR hybrids~\cite{Ajith:2012az}, 
which are being used to study the efficacy of 
search and parameter-estimation codes, and the Numerical-Relativity--Analytical-Relativity (NRAR) 
collaboration has performed an extensive study of the accuracy of a collection of 27 NR 
waveforms~\cite{Hinder:2013oqa}. The NINJA study is currently limited to non-precessing binaries, 
and the quantification of waveform accuracies, and their requirements for modelling, remain problematic,
but recent progress in numerical simulations is certainly comparable to the progress in modelling. 

Beyond modelling issues, we also lack a search strategy for generic binaries, besides the first-approximation
expectation that non-precessing-binary models will be sufficient to capture a large number of generic
sources~\cite{Harry:2013tca}. By definition a search that employed full generic-binary models would 
locate a wider range of signals,
but it is possible that the increased false-alarm rate incurred by including more parameters in the 
search model would outweigh the better agreement between the generic-binary sources and the 
search templates. 

In the long term (after $\sim$2025), if third-generation and space-based detectors come 
online~\cite{Punturo:2010zz,Seoane:2013qna},
then the situation may be quite different. As parameter degeneracies become less important and more
accurate measurements become possible, the accuracy requirements of the waveform models will
increase, and higher harmonics will be essential. It may also be possible at that stage to perform stringent
tests of the general theory of relativity~\cite{lrr-2013-9}; such tests will require far greater waveform precision 
than astrophysical measurements, although the details are yet to be clarified.

\begin{acknowledgements}
Thanks to 
Alejandro Boh\'e,
Stephen Fairhurst,
Maxime Fays, 
Sascha Husa,
Frank Ohme, 
Francesco Pannarale, 
B. Sathyaprakash 
and
Patricia Schmidt
for useful discussions and/or comments on the manuscript. This work was 
supported by STFC grants ST/H008438/1 and ST/I001085/1.
\end{acknowledgements}

\bibliographystyle{h-physrev}
\bibliography{review.bib}

\begin{thebibliography}{10}

\bibitem{Harry:2010zz}
LIGO Scientific Collaboration, G.~M. Harry,
\newblock Class.Quant.Grav. {\bf 27}, 084006 (2010).

\bibitem{aVIRGO}
{The Virgo Collaboration}, F.~Acernese {\em et~al.},
\newblock {Advanced Virgo Baseline Design}, 2009,
\newblock {[Virgo Techincal Document VIR-0027A-09]}.

\bibitem{Aasi:2013wya}
LIGO Scientific Collaboration, Virgo Collaboration, J.~Aasi {\em et~al.},
\newblock {Prospects for Localization of Gravitational Wave Transients by the
  Advanced LIGO and Advanced Virgo Observatories},
\newblock 2013.

\bibitem{Abadie:2010cf}
LIGO Scientific Collaboration, Virgo Collaboration, J.~Abadie {\em et~al.},
\newblock Class.Quant.Grav. {\bf 27}, 173001 (2010), 1003.2480.

\bibitem{Blanchet:2013haa}
L.~Blanchet,
\newblock (2013), 1310.1528.

\bibitem{Centrella:2010mx}
J.~Centrella, J.~G. Baker, B.~J. Kelly, and J.~R. van Meter,
\newblock Rev.Mod.Phys. {\bf 82}, 3069 (2010), 1010.5260.

\bibitem{Ajith:2009bn}
P.~Ajith {\em et~al.},
\newblock Phys.Rev.Lett. {\bf 106}, 241101 (2011), 0909.2867.

\bibitem{Pan:2009wj}
Y.~Pan {\em et~al.},
\newblock Phys.Rev. {\bf D81}, 084041 (2010), 0912.3466.

\bibitem{Santamaria:2010yb}
L.~Santamaria {\em et~al.},
\newblock Phys.Rev. {\bf D82}, 064016 (2010), 1005.3306.

\bibitem{Taracchini:2012ig}
A.~Taracchini {\em et~al.},
\newblock Phys.Rev. {\bf D86}, 024011 (2012), 1202.0790.

\bibitem{Taracchini:2013rva}
A.~Taracchini {\em et~al.},
\newblock (2013), 1311.2544.

\bibitem{Barker:1975ae}
B.~Barker and R.~O'Connell,
\newblock Phys.Rev. {\bf D12}, 329 (1975).

\bibitem{Barker:1979}
B.~Barker and R.~O'Connell,
\newblock Gen. Rel. Grav. {\bf 11}, 149 (1979).

\bibitem{Damour87}
T.~Damour,
\newblock ???,
\newblock in {\em 300 Years of Gravitation}, edited by S.~Hawking and
  W.~Israel, Cambridge University Press, 1987.

\bibitem{Apostolatos:1994mx}
T.~A. Apostolatos, C.~Cutler, G.~J. Sussman, and K.~S. Thorne,
\newblock Phys. Rev. {\bf D49}, 6274 (1994).

\bibitem{Kidder:1995zr}
L.~E. Kidder,
\newblock Phys. Rev. {\bf D52}, 821 (1995), gr-qc/9506022.

\bibitem{Peters:1964zz}
P.~Peters,
\newblock Phys.Rev. {\bf 136}, B1224 (1964).

\bibitem{Buonanno:2002fy}
A.~Buonanno, Y.-b. Chen, and M.~Vallisneri,
\newblock Phys.Rev. {\bf D67}, 104025 (2003), gr-qc/0211087.

\bibitem{Blanchet:2006gy}
L.~Blanchet, A.~Buonanno, and G.~Faye,
\newblock Phys.Rev. {\bf D74}, 104034 (2006), gr-qc/0605140.

\bibitem{Colaboration:2011np}
LIGO Collaboration, Virgo Collaboration, J.~Abadie {\em et~al.},
\newblock Phys.Rev. {\bf D85}, 082002 (2012), 1111.7314.

\bibitem{Aasi:2012rja}
LIGO Scientific Collaboration, Virgo Collaboration, J.~Aasi {\em et~al.},
\newblock Phys.Rev. {\bf D87}, 022002 (2013), 1209.6533.

\bibitem{Jaranowski:2013lca}
P.~Jaranowski and G.~Schäfer,
\newblock Phys.Rev. {\bf D87}, 081503 (2013), 1303.3225.

\bibitem{Hartung:2010jg}
J.~Hartung and J.~Steinhoff,
\newblock Phys.Rev. {\bf D83}, 044008 (2011), 1011.1179.

\bibitem{Marsat:2012fn}
S.~Marsat, A.~Boh\'e, G.~Faye, and L.~Blanchet,
\newblock Class.Quantum Grav. {\bf 30}, 055007 (2013), 1210.4143.

\bibitem{Levi:2011eq}
M.~Levi,
\newblock Phys.Rev. {\bf D85}, 064043 (2012), 1107.4322.

\bibitem{Hartung:2011ea}
J.~Hartung and J.~Steinhoff,
\newblock Annalen Phys. {\bf 523}, 919 (2011), 1107.4294.

\bibitem{Bohe:2013cla}
A.~Boh\'e, S.~Marsat, and L.~Blanchet,
\newblock Class.Quant.Grav. {\bf 30}, 135009 (2013), 1303.7412.

\bibitem{Marsat:2013caa}
S.~Marsat, A.~Bohe, L.~Blanchet, and A.~Buonanno,
\newblock (2013), 1307.6793.

\bibitem{Gergely:1999pd}
L.~A. Gergely,
\newblock Phys.Rev. {\bf D61}, 024035 (2000), gr-qc/9911082.

\bibitem{Faye:2012xt}
G.~Faye, S.~Marsat, L.~Blanchet, and B.~R. Iyer,
\newblock (2012), 1210.2339.

\bibitem{Arun:2008kb}
K.~Arun, A.~Buonanno, G.~Faye, and E.~Ochsner,
\newblock Phys.Rev. {\bf D79}, 104023 (2009), 0810.5336.

\bibitem{Buonanno:2009zt}
A.~Buonanno, B.~Iyer, E.~Ochsner, Y.~Pan, and B.~Sathyaprakash,
\newblock Phys.Rev. {\bf D80}, 084043 (2009), 0907.0700.

\bibitem{Nitz:2013mxa}
A.~H. Nitz {\em et~al.},
\newblock (2013), 1307.1757.

\bibitem{Brown:2012qf}
D.~A. Brown, I.~Harry, A.~Lundgren, and A.~H. Nitz,
\newblock Phys.Rev. {\bf D86}, 084017 (2012), 1207.6406.

\bibitem{Hannam:2007ik}
M.~Hannam, S.~Husa, U.~Sperhake, B.~Bruegmann, and J.~A. Gonzalez,
\newblock Phys.Rev. {\bf D77}, 044020 (2008), 0706.1305.

\bibitem{Boyle:2007ft}
M.~Boyle {\em et~al.},
\newblock Phys.Rev. {\bf D76}, 124038 (2007), 0710.0158.

\bibitem{Hannam:2007wf}
M.~Hannam, S.~Husa, B.~Bruegmann, and A.~Gopakumar,
\newblock Phys.Rev. {\bf D78}, 104007 (2008), 0712.3787.

\bibitem{Hannam:2010ec}
M.~Hannam, S.~Husa, F.~Ohme, D.~Muller, and B.~Bruegmann,
\newblock Phys.Rev. {\bf D82}, 124008 (2010), 1007.4789.

\bibitem{MacDonald:2012mp}
I.~MacDonald {\em et~al.},
\newblock Phys.Rev. {\bf D87}, 024009 (2013), 1210.3007.

\bibitem{Buonanno:1998gg}
A.~Buonanno and T.~Damour,
\newblock Phys.Rev. {\bf D59}, 084006 (1999), gr-qc/9811091.

\bibitem{Buonanno:2000ef}
A.~Buonanno and T.~Damour,
\newblock Phys.Rev. {\bf D62}, 064015 (2000), gr-qc/0001013.

\bibitem{Damour:2001tu}
T.~Damour,
\newblock Phys.Rev. {\bf D64}, 124013 (2001), gr-qc/0103018.

\bibitem{Buonanno:2005xu}
A.~Buonanno, Y.~Chen, and T.~Damour,
\newblock Phys.Rev. {\bf D74}, 104005 (2006), gr-qc/0508067.

\bibitem{Pretorius:2005gq}
F.~Pretorius,
\newblock Phys. Rev. Lett. {\bf 95}, 121101 (2005), gr-qc/0507014.

\bibitem{Campanelli:2005dd}
M.~Campanelli, C.~O. Lousto, P.~Marronetti, and Y.~Zlochower,
\newblock Phys. Rev. Lett. {\bf 96}, 111101 (2006), gr-qc/0511048.

\bibitem{Baker:2005vv}
J.~G. Baker, J.~Centrella, D.-I. Choi, M.~Koppitz, and J.~van Meter,
\newblock Phys. Rev. Lett. {\bf 96}, 111102 (2006), gr-qc/0511103.

\bibitem{Ajith:2007qp}
P.~Ajith {\em et~al.},
\newblock Class.Quant.Grav. {\bf 24}, S689 (2007), 0704.3764.

\bibitem{Buonanno:2006ui}
A.~Buonanno, G.~B. Cook, and F.~Pretorius,
\newblock Phys.Rev. {\bf D75}, 124018 (2007), gr-qc/0610122.

\bibitem{Buonanno:2007sv}
A.~Buonanno, L.~E. Kidder, and L.~Lehner,
\newblock Phys.Rev. {\bf D77}, 026004 (2008), 0709.3839.

\bibitem{Hemberger:2013hsa}
D.~A. Hemberger {\em et~al.},
\newblock Phys.Rev. {\bf D88}, 064014 (2013), 1305.5991.

\bibitem{Barausse:2009uz}
E.~Barausse and L.~Rezzolla,
\newblock Astrophys.J. {\bf 704}, L40 (2009).

\bibitem{Poisson:1995ef}
E.~Poisson and C.~M. Will,
\newblock Phys.Rev. {\bf D52}, 848 (1995), gr-qc/9502040.

\bibitem{Ajith:2011ec}
P.~Ajith,
\newblock Phys.Rev. {\bf D84}, 084037 (2011), 1107.1267.

\bibitem{Ajith:2007kx}
P.~Ajith {\em et~al.},
\newblock Phys.Rev. {\bf D77}, 104017 (2008), 0710.2335.

\bibitem{Buonanno:2007pf}
A.~Buonanno {\em et~al.},
\newblock Phys.Rev. {\bf D76}, 104049 (2007), 0706.3732.

\bibitem{Damour:2008gu}
T.~Damour, B.~R. Iyer, and A.~Nagar,
\newblock Phys.Rev. {\bf D79}, 064004 (2009), 0811.2069.

\bibitem{Pan:2010hz}
Y.~Pan, A.~Buonanno, R.~Fujita, E.~Racine, and H.~Tagoshi,
\newblock Phys.Rev. {\bf D83}, 064003 (2011), 1006.0431.

\bibitem{Damour:2012ky}
T.~Damour, A.~Nagar, and S.~Bernuzzi,
\newblock Phys.Rev. {\bf D87}, 084035 (2013), 1212.4357.

\bibitem{Pan:2011gk}
Y.~Pan {\em et~al.},
\newblock Phys.Rev. {\bf D84}, 124052 (2011), 1106.1021.

\bibitem{Barausse:2009xi}
E.~Barausse and A.~Buonanno,
\newblock Phys.Rev. {\bf D81}, 084024 (2010), 0912.3517.

\bibitem{Purrer:2013ojf}
M.~P{\"u}rrer, M.~Hannam, P.~Ajith, and S.~Husa,
\newblock Phys.Rev. {\bf D88}, 064007 (2013), 1306.2320.

\bibitem{Gonzalez:2007hi}
J.~Gonzalez, M.~Hannam, U.~Sperhake, B.~Bruegmann, and S.~Husa,
\newblock Phys.Rev.Lett. {\bf 98}, 231101 (2007), gr-qc/0702052.

\bibitem{Campanelli:2007cga}
M.~Campanelli, C.~O. Lousto, Y.~Zlochower, and D.~Merritt,
\newblock Phys.Rev.Lett. {\bf 98}, 231102 (2007), gr-qc/0702133.

\bibitem{Brugmann:2007zj}
B.~Bruegmann, J.~A. Gonzalez, M.~Hannam, S.~Husa, and U.~Sperhake,
\newblock Phys.Rev. {\bf D77}, 124047 (2008), 0707.0135.

\bibitem{Sturani:2010yv}
R.~Sturani {\em et~al.},
\newblock J.Phys.Conf.Ser. {\bf 243}, 012007 (2010), 1005.0551.

\bibitem{Apostolatos:1995pj}
T.~Apostolatos,
\newblock Phys.Rev. {\bf D52}, 605 (1995).

\bibitem{Pan:2003qt}
Y.~Pan, A.~Buonanno, Y.-b. Chen, and M.~Vallisneri,
\newblock Phys.Rev. {\bf D69}, 104017 (2004), gr-qc/0310034.

\bibitem{Schmidt:2010it}
P.~Schmidt, M.~Hannam, S.~Husa, and P.~Ajith,
\newblock Phys.Rev. {\bf D84}, 024046 (2011), 1012.2879.

\bibitem{O'Shaughnessy:2011fx}
R.~O'Shaughnessy, B.~Vaishnav, J.~Healy, Z.~Meeks, and D.~Shoemaker,
\newblock Phys.Rev. {\bf D84}, 124002 (2011), 1109.5224.

\bibitem{Boyle:2011gg}
M.~Boyle, R.~Owen, and H.~P. Pfeiffer,
\newblock Phys.Rev. {\bf D84}, 124011 (2011), 1110.2965.

\bibitem{Schmidt:2012rh}
P.~Schmidt, M.~Hannam, and S.~Husa,
\newblock Phys.Rev. {\bf D86}, 104063 (2012), 1207.3088.

\bibitem{Lundgren:2013jla}
A.~Lundgren and R.~O'Shaughnessy,
\newblock (2013), 1304.3332.

\bibitem{Pekowsky:2013ska}
L.~Pekowsky, R.~O'Shaughnessy, J.~Healy, and D.~Shoemaker,
\newblock Phys.Rev. {\bf D88}, 024040 (2013), 1304.3176.

\bibitem{Hannam:2013oca}
M.~Hannam {\em et~al.},
\newblock (2013), 1308.3271.

\bibitem{Schmidt:2013aa}
P.~Schmidt, M.~Hannam, and F.~Ohme,
\newblock (2013),
\newblock In preparation.

\bibitem{Pan:2013rra}
Y.~Pan {\em et~al.},
\newblock (2013), 1307.6232.

\bibitem{O'Shaughnessy:2012ay}
R.~O'Shaughnessy, L.~London, J.~Healy, and D.~Shoemaker,
\newblock Phys.Rev. {\bf D87}, 044038 (2013), 1209.3712.

\bibitem{MacDonald:2011ne}
I.~MacDonald, S.~Nissanke, and H.~P. Pfeiffer,
\newblock Class.Quant.Grav. {\bf 28}, 134002 (2011), 1102.5128.

\bibitem{Damour:2010zb}
T.~Damour, A.~Nagar, and M.~Trias,
\newblock Phys.Rev. {\bf D83}, 024006 (2011), 1009.5998.

\bibitem{Boyle:2011dy}
M.~Boyle,
\newblock Phys.Rev. {\bf D84}, 064013 (2011), 1103.5088.

\bibitem{Hannam:2010ky}
M.~Hannam, S.~Husa, F.~Ohme, and P.~Ajith,
\newblock Phys.Rev. {\bf D82}, 124052 (2010), 1008.2961.

\bibitem{Ohme:2011zm}
F.~Ohme, M.~Hannam, and S.~Husa,
\newblock Phys.Rev. {\bf D84}, 064029 (2011), 1107.0996.

\bibitem{Capano:2013raa}
C.~Capano, Y.~Pan, and A.~Buonanno,
\newblock (2013), 1311.1286.

\bibitem{Cutler:1994ys}
C.~Cutler and E.~E. Flanagan,
\newblock Phys.Rev. {\bf D49}, 2658 (1994), gr-qc/9402014.

\bibitem{Baird:2012cu}
E.~Baird, S.~Fairhurst, M.~Hannam, and P.~Murphy,
\newblock Phys.Rev. {\bf D87}, 024035 (2013), 1211.0546.

\bibitem{Hannam:2013uu}
M.~Hannam, D.~A. Brown, S.~Fairhurst, C.~L. Fryer, and I.~W. Harry,
\newblock Astrophys.J. {\bf 766}, L14 (2013), 1301.5616.

\bibitem{Ohme:2013nsa}
F.~Ohme, A.~B. Nielsen, D.~Keppel, and A.~Lundgren,
\newblock Phys.Rev. {\bf D88}, 042002 (2013), 1304.7017.

\bibitem{Mroue:2013xna}
A.~H. Mroue {\em et~al.},
\newblock (2013), 1304.6077.

\bibitem{Aylott:2009ya}
B.~Aylott {\em et~al.},
\newblock Class.Quant.Grav. {\bf 26}, 165008 (2009), 0901.4399.

\bibitem{Ajith:2012az}
P.~Ajith {\em et~al.},
\newblock Class.Quant.Grav. {\bf 29}, 124001 (2012), 1201.5319.

\bibitem{Hinder:2013oqa}
I.~Hinder {\em et~al.},
\newblock Class.Quant.Grav. {\bf 31}, 025012 (2013), 1307.5307.

\bibitem{Harry:2013tca}
I.~Harry {\em et~al.},
\newblock (2013), 1307.3562.

\bibitem{Punturo:2010zz}
M.~Punturo {\em et~al.},
\newblock Class.Quant.Grav. {\bf 27}, 194002 (2010).

\bibitem{Seoane:2013qna}
eLISA Collaboration, P.~A. Seoane {\em et~al.},
\newblock (2013), 1305.5720.

\bibitem{lrr-2013-9}
N.~Yunes and X.~Siemens,
\newblock Living Reviews in Relativity {\bf 16} (2013).

\end{thebibliography}

\end{document}